\documentclass[fleqn,usenatbib]{mnras}
 \usepackage{graphicx}
 \usepackage{amsmath}
 \usepackage{amssymb}
 \usepackage[T1]{fontenc}
 \usepackage{ae,aecompl}
 \usepackage{txfonts}
 \usepackage{enumerate}
 \title[Structures beyond the Kuiper Cliff]
       {Past the outer rim, into the unknown: structures beyond the Kuiper Cliff}

 \author[C. de la Fuente Marcos and R. de la Fuente Marcos]
        {C.~de~la~Fuente~Marcos$^{1}$\thanks{E-mail: nbplanet@ucm.es}
         and
         R. de la Fuente Marcos$^{2}$ \\
         $^1$Universidad Complutense de Madrid,
             Ciudad Universitaria, E-28040 Madrid, Spain \\
         $^2$AEGORA Research Group,
             Facultad de Ciencias Matem\'aticas,
             Universidad Complutense de Madrid,
             Ciudad Universitaria, E-28040 Madrid, Spain}
 \date{Accepted 2023 September 01. 
       Received 2023 August 31; 
       in original form 2023 August 11}
 \pubyear{2023}
 
\begin{document}
  \label{firstpage}
  \pagerange{\pageref{firstpage}--\pageref{lastpage}}
  \maketitle
%
%
  \begin{abstract}
     Although the present-day orbital distribution of minor bodies that go 
     around the Sun between the orbit of Neptune and the Kuiper Cliff is well 
     understood, past $\sim$50~au from the Sun, our vision gets blurred as 
     objects become fainter and fainter and their orbital periods span 
     several centuries. Deep imaging using the largest telescopes can 
     overcome the first issue but the problems derived from the second one 
     are better addressed using data analysis techniques. Here, we make use 
     of the heliocentric range and range-rate of the known Kuiper belt 
     objects and their uncertainties to identify structures in orbital 
     parameter space beyond the Kuiper Cliff. The distribution in 
     heliocentric range there closely resembles that of the outer main 
     asteroid belt with a gap at $\sim$70~au that may signal the existence 
     of a dynamical analogue of the Jupiter family comets. Outliers in the 
     distribution of mutual nodal distances suggest that a massive perturber 
     is present beyond the heliopause.
  \end{abstract}

  \begin{keywords}
     methods: data analysis -- celestial mechanics -- 
     minor planets, asteroids: general -- Kuiper belt: general.
  \end{keywords}

  \section{Introduction}
     The Solar system beyond Neptune was a great unknown when (134340) Pluto 1930~BM was discovered by C.~W. Tombaugh 
     \citep{1930PASP...42..105A}. It was soon suggested that a population of bodies in Pluto-like orbits existed beyond 
     Neptune \citep{1930ASPL....1..121L} and this hypothesis was independently explored by several authors (see e.g. 
     \citealt{1943JBAA...53..181E,1949MNRAS.109..600E,1951astr.conf..357K,1962Icar....1...13C,1978M&P....18....5C,
     1964PNAS...51..711W,1972IAUS...45..401W,1980MNRAS.192..481F}). The credibility of this conjecture was confirmed 
     numerically by \citet{1988ApJ...328L..69D}, but the observational proof had to wait until 1992 when the second 
     member of this population, (15760) Albion 1992~QB$_{1}$ was found \citep{1992IAUC.5611....1J,1993Natur.362..730J}.

     Three decades and thousands of discoveries later, our understanding of the organization of the Solar system beyond 
     the orbit of Neptune is satisfactory when we consider the classical Edgeworth-Kuiper belt (see e.g. 
     \citealt{2020tnss.book...25M}). But beyond $\sim$50~au from the Sun, there is a statistically significant decrease 
     in the number density of observed objects (see e.g. \citealt{1999AJ....118.1411C}), the so-called Kuiper Cliff, 
     whose origin remains elusive, although its reality is supported by the data (see e.g. 
     \citealt{2021ARA&A..59..203G}). Here, we present an alternative approach for identifying structures beyond the 
     Kuiper Cliff that instead of orbits uses the heliocentric range, $r$, and range-rate, $\dot{r}$, of the known 
     Kuiper belt objects together with their respective uncertainties, $\sigma_r$ and $\sigma_{\dot{r}}$, from a given 
     epoch. This Letter is organized as follows. In Section~2, we present the data and data analysis tools used. The 
     latest data available on the edge of the known Solar system are reviewed in Section~3. The convenience of the 
     analysis of heliocentric range and range-rate data is discussed and assessed in Section~4. Section~5 focuses on 
     what can be learned from the best sample. Our results are discussed in Section~6 and our conclusions are summarized 
     in Section~7.

  \section{Data and tools}
     In this work, we use ephemerides computed by Jet Propulsion Laboratory's (JPL) Horizons on-line solar system data 
     and ephemeris computation service\footnote{\href{https://ssd.jpl.nasa.gov/?horizons} 
     {https://ssd.jpl.nasa.gov/?horizons}} \citep{2015IAUGA..2256293G} that utilises the new DE440/441 general-purpose 
     planetary solution \citep{2021AJ....161..105P}. Data queries were made via the \textsc{Python} package 
     \textsc{Astroquery} \citep{2019AJ....157...98G}. Our input data sample was retrieved from JPL's Small-Body Database 
     (SBDB).\footnote{\href{https://ssd.jpl.nasa.gov/sbdb.cgi}{https://ssd.jpl.nasa.gov/sbdb.cgi}} It includes all the 
     4474 objects (as of 30-Aug-2023) in the trans-Neptunian object orbit class (semimajor axis, $a>30.1$~au). The data 
     collected are referred to the standard epoch JD 2460200.5 (2023-Sep-13.0) TDB (Barycentric Dynamical Time, J2000.0 
     ecliptic and equinox). Statistical analyses were carried out using \textsc{NumPy} \citep{2011CSE....13b..22V, 
     2020Natur.585..357H} and \textsc{Astropy} \citep{2013A&A...558A..33A,2018AJ....156..123A}, and visualized using the 
     \textsc{Matplotlib} library \citep{2007CSE.....9...90H}. Histograms use a bin width computed by applying the 
     Freedman-Diaconis rule \citep{FD81}, i.e. $2\ {\rm IQR}\ n^{-1/3}$, where IQR is the interquartile range and $n$ is 
     the number of data points. The distant nature of relevant objects was confirmed statistically using the software 
     discussed by \citet{2000AJ....120.3323B}. 

  \section{The edge of the known Solar system}
     The local interstellar medium starts at the heliopause, $\sim$120~au from the Sun (see e.g. 
     \citealt{2023ApJ...951...71K}), where there is a sharp transition from a hot and tenuous solar wind to a colder, 
     denser interstellar plasma. Until recently, the Solar system beyond 100~au from the Sun was exclusively studied 
     using data provided by interplanetary probes like Voyager 1 and 2 (see e.g. \citealt{2020ApJ...900L...1K,
     2021ApJ...911...61B,2021ApJ...921...62G,2022A&A...658L..12M}). The 100~au barrier was finally broken in 2018 and 
     four minor bodies have so far been found around the heliopause and beyond (see Table~\ref{beyond100au}). Distant 
     Solar system bodies are identified through their motion against the background sky. Objects detected at nearly 
     100~au from the Sun show a maximum observable apparent motion of about 33{\arcsec}~d$^{-1}$ and those located 
     farther away have progressively lower apparent rates of motion (all due to parallax, see e.g. 
     \citealt{1999AREPS..27..287J}). 
%
%
     \begin{table}
        \centering
        \fontsize{8}{12pt}\selectfont
        \tabcolsep 0.15truecm
        \caption{\label{beyond100au}Known trans-Neptunian objects currently located beyond 100~au from the Sun: 
                 2018~VG$_{18}$ was discovered in November 2018 at 123~au \citep{2018MPEC....Y...14S,
                 2020MPEC....A..128S}, 2018~AG$_{37}$ was first observed in January 2018 at 132~au 
                 \citep{2021MPEC....C..187S}, 2020~BE$_{102}$ was found at 111~au in January 2020 
                 \citep{2022MPEC....K..172S}, and 2020~MK$_{53}$ was identified in June 2020 at 156~au. Data sources: 
                 JPL's SBDB and Horizons.
                }
        \begin{tabular}{lcccccc}
           \hline
            Object          & $r$    & 3 $\sigma_r$ & $\dot{r}$     & 3 $\sigma_{\dot{r}}$ & $a$   & $\sigma_a$ \\
                            & (au)   & (au)         & (km~s$^{-1}$) & (km~s$^{-1}$)        & (au)  & (au)       \\
           \hline
            2018~AG$_{37}$  & 132    & 8            & $-$0.1        & 9.1                  & 80    & 4          \\
            2018~VG$_{18}$  & 123.73 & 0.05         &    0.28       & 0.02                 & 81.96 & 0.03       \\
            2020~BE$_{102}$ & 110.7  & 0.9          & $-$0.7        & 1.7                  & 75    & 5          \\
            2020~MK$_{53}$  & 160    & 66917        & $-$0.04       & 105589               & 111   & 21051      \\
           \hline
        \end{tabular}
     \end{table}
%
%

     The current record holder is 2020~MK$_{53}$ found by New Horizons KBO Search-Subaru \citep{2022DPS....5450109P}. 
     For this object, the software discussed in \citet{2000AJ....120.3323B} applied to the available data (six 
     observations spanning three days at $r'$=26~mag) gives a barycentric distance of 156$\pm$4~au and the latest JPL's 
     Horizons ephemerides give 160$\pm$22306~au (2460200.5 JD TDB) as the uncertainty grows exponentially over time for 
     short data arcs. In general and for distant objects, the few observations in the short arc constrain the distance 
     from the observer (topocentric range) and the radial velocity (topocentric range-rate) but for objects as distant 
     as the heliopause, topocentric, geocentric, heliocentric and barycentric ranges and range-rates are affected by 
     very similar uncertainties.

     \citet{2021AJ....162...39O} studied the orbital distribution of the objects that populate the outer Solar system 
     and argued that there is a gap in perihelion distance that separates two populations, the extreme trans-Neptunian 
     objects (ETNOs, \citealt{2014Natur.507..471T}) and the inner Oort Cloud objects (IOCOs, 
     \citealt{1981AJ.....86.1730H,2001AJ....121.2253L}). These authors predict that such a gap may be located at 
     50--65~au but their conclusions are based on a very small sample and numerical simulations. This is consistent with
     early results discussed by \citet{2023LPICo2806.2361F}, see their fig.~2.  

  \section{Orbit versus range and range-rate}
     The orbit determinations of distant objects based on data arcs shorter than about a year are very unreliable and 
     their associated uncertainties could be very large (often $>50$ per cent). However, their geocentric, heliocentric 
     or barycentric distances estimated for an epoch chosen between the dates of their first and last observation could 
     be uncertain by a few percent as pointed out above for the extreme case of 2020~MK$_{53}$. This relatively small 
     uncertainty grows rapidly for predictions outside the timeframe defined by the available data arc. 

\begin{figure}
       \centering
        \includegraphics[width=\linewidth]{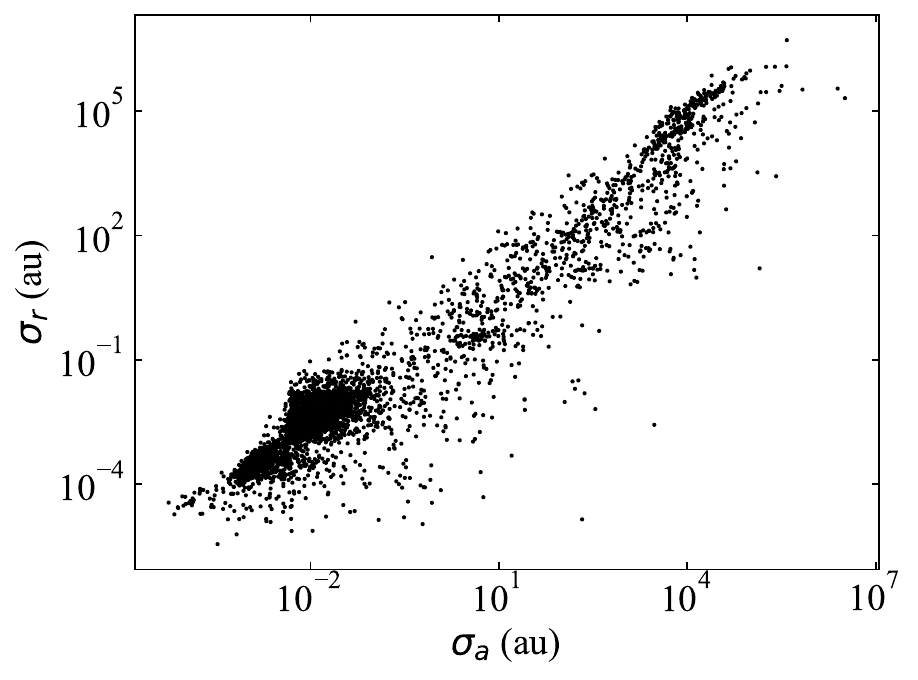}
        \includegraphics[width=\linewidth]{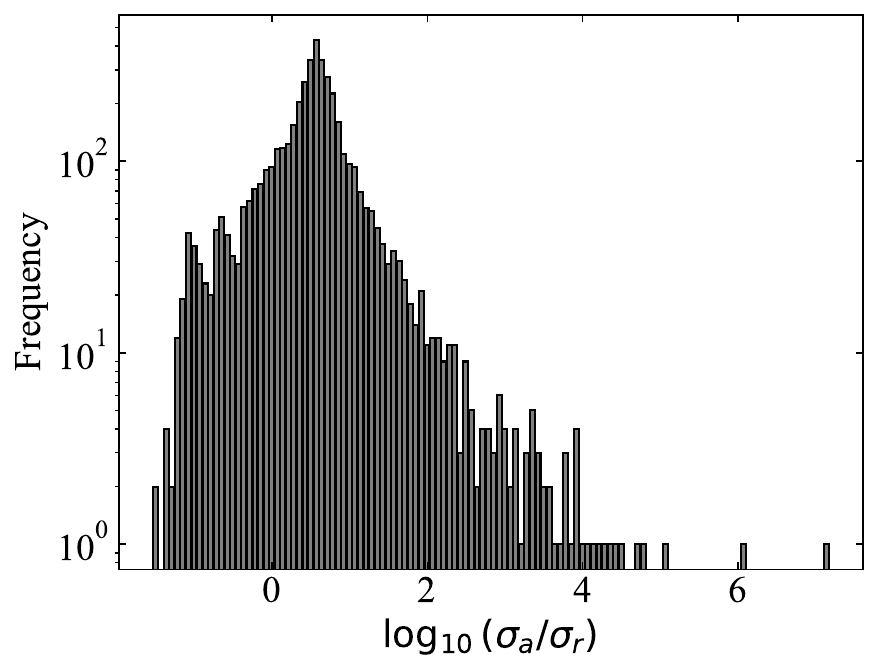}
        \caption{Uncertainty in heliocentric range as a function of the uncertainty in the value of the semimajor axis 
                 (top panel) for the input sample. Note the strong correlation. Histogram of the logarithm 
                 of the ratio of uncertainties (bottom panel), the median value is 0.5461, the quartiles are 
                 0.2035 and 0.7935. Data sources: JPL's SBDB and Horizons. 
                }
        \label{rangevsa}
     \end{figure}

     An object located at 100~au from the Sun moving with radial velocity under 1~km~s$^{-1}$ can travel $<0.63$~au 
     in three years, although the nominal uncertainty of $r$ could be in the hundreds or thousands of au. This means 
     that the predicted nominal heliocentric range is probably not too far from the real value even for objects with 
     relatively short data arcs. Our working hypothesis, that heliocentric range and range-rate can provide usable 
     information when orbit determinations cannot, can be tested using the available data. Figure~\ref{rangevsa} shows 
     an example for the particular case of the semimajor axis, $a$, and its uncertainty, $\sigma_{a}$, and confirms that 
     our hypothesis is correct as $\sigma_{a}>\sigma_{r}$, even for an epoch (JD 2460200.5 TDB) some years outside the 
     available data arc. Therefore, $r$ and $\dot{r}$ can be used with confidence to investigate the presence of 
     structures beyond the Kuiper Cliff.
%
%
%
%

\begin{figure*}
       \centering
        \includegraphics[width=0.33\linewidth]{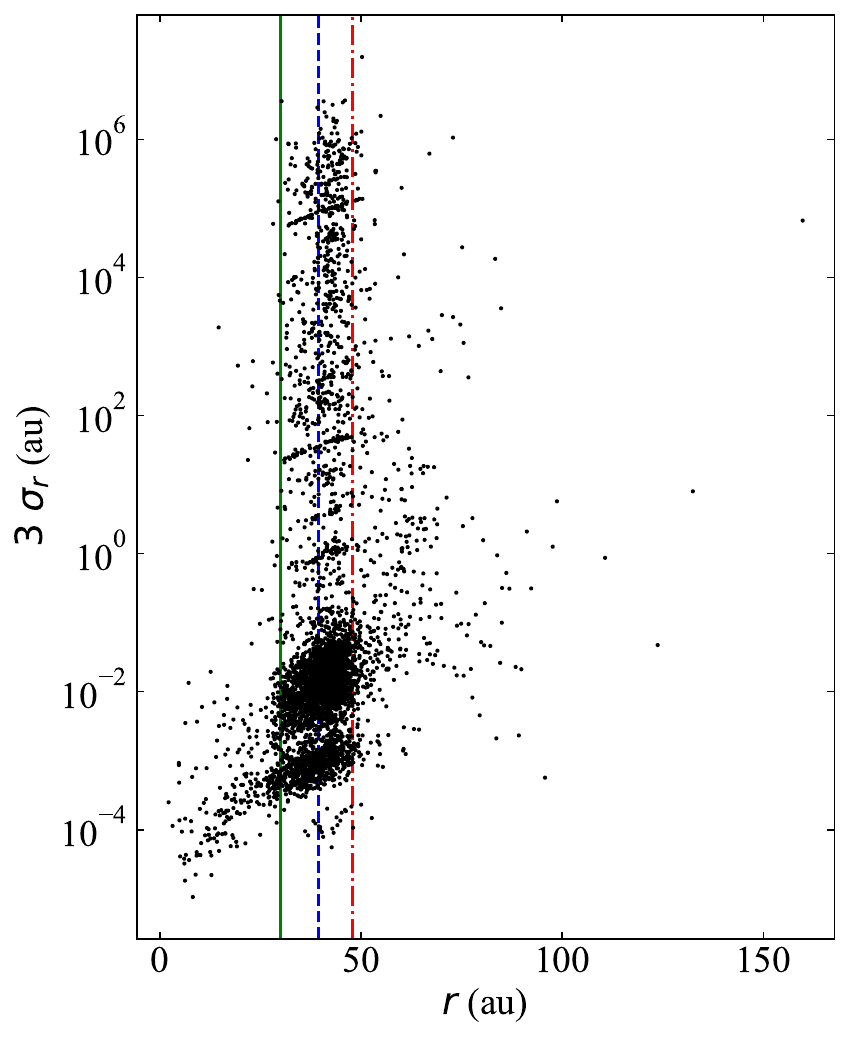}
        \includegraphics[width=0.33\linewidth]{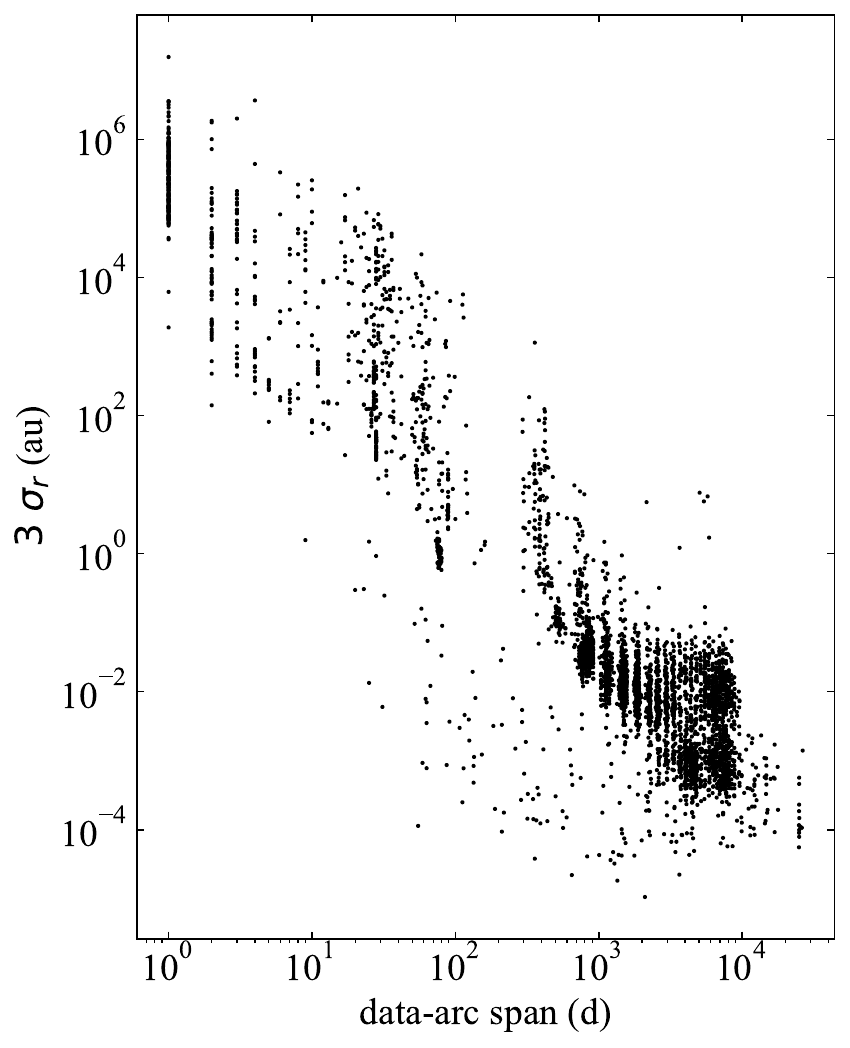}
        \includegraphics[width=0.33\linewidth]{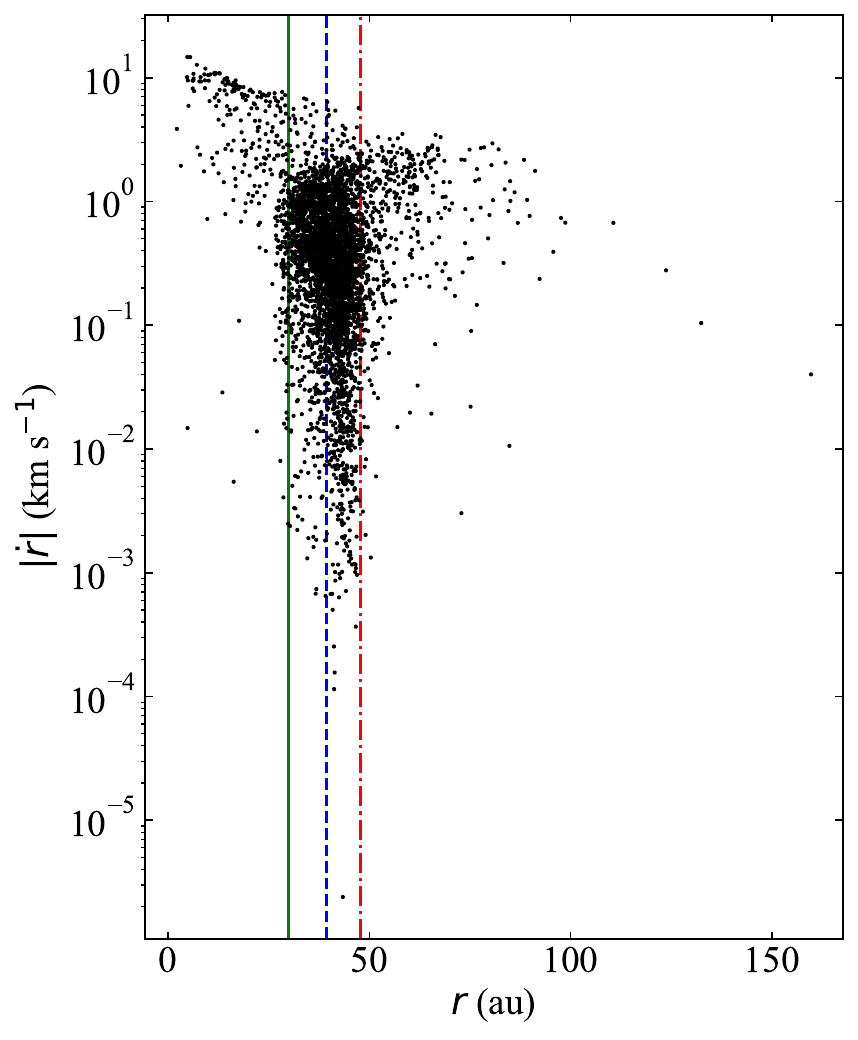}
        \caption{Heliocentric range and range-rate for the known members of the trans-Neptunian object orbit class 
                 ($a>30.1$~au), 4474 as of 30-Aug-2023. The left-hand side panel shows the uncertainty in heliocentric 
                 range as a function of the range. The central panel shows that uncertainty in heliocentric range as a 
                 function of the data-arc span. The right-hand side panel shows the absolute value of the heliocentric 
                 range-rate as a function of the range with low values of the range-rate indicating proximity to 
                 perihelion or aphelion. The location of the 1:1 mean-motion resonance with Neptune at 30.0~au is 
                 displayed as a green solid vertical line, the 2:3 resonance at 39.4~au as blue dashed and 
                 the 1:2 resonance at 47.8~au as red dot-dashed signalling the Kuiper Cliff. Bands observed in the first 
                 two panels are likely the result of surveys having different observing cadences and recovery 
                 strategies. Data sources: JPL's SBDB and Horizons.
                }
        \label{uncertainty}
     \end{figure*}

     The relevant ephemerides of the input sample are plotted in Fig.~\ref{uncertainty}. The left-hand side panel shows 
     how the uncertainty in heliocentric range changes as a function of the range. In addition to the conspicuous Kuiper 
     Cliff (the 1:2 mean-motion resonance with Neptune at 47.8~au, see \citealt{2001ApJ...554L..95T}), we observe bands 
     of objects (somewhat inclined with respect to the $x$-axis) with different levels of predicted nominal uncertainty 
     for those minor bodies that revolve around the Sun between the orbit of Neptune and the Cliff. Beyond this region, 
     we observe less data points with larger uncertainties. 

     The origin of the uncertainty bands for radial distances in the range 30--50~au in the left-hand side panel becomes 
     apparent in the central panel, where the uncertainty in heliocentric range is plotted as a function of the data-arc 
     span. Objects with orbit determinations based on short data arcs have larger uncertainties in range. The bands 
     (vertical in the central panel and slightly inclined in the left-hand side panel) are likely the result of the 
     different observing cadences and recovery strategies of the various surveys contributing data. 

     Considering our previous arguments on the relatively small errors associated to range and range-rate for times 
     close or within the data arc, the figure suggests that objects with very uncertain radial distances (top of the 
     left-hand side panel) tend to keep their nominal values within the boundaries 30.0 au to 47.8~au as their 
     observational data arcs grow longer and the uncertainties decrease. For a new discovery, the usual trend, derived 
     from the left-hand side and the central panels of Fig.~\ref{uncertainty}, is to move vertically downwards in the 
     diagram that shows the uncertainty in heliocentric range as a function of the value of the range. If this strictly 
     observational tendency is valid in the interval (30, 50)~au, there are no reasons to believe that for heliocentric 
     distances $>50$~au the evolution has to be different. 

     The right-hand side panel of Fig.~\ref{uncertainty} shows the absolute values of the associated radial velocities 
     as a function of the radial distance. Such values get close to zero at aphelion and perihelion. In a Keplerian disc 
     of particles, a gap is expected to appear at a radial distance $r_{\rm g}$ if a significant number of particles 
     reaches apocentre at $r<r_{\rm g}$ and another significant fraction reaches pericentre at $r>r_{\rm g}$. Gaps also 
     appear linked to peculiar populations in eccentric orbits like that of asteroids dynamically similar to Jupiter 
     Family Comets (JFCs, see e.g. \citealt{2016A&A...589A.111G}).
%
%
%
%

  \section{Results based on the best sample}
\begin{figure*}
       \centering
        \includegraphics[width=0.33\linewidth]{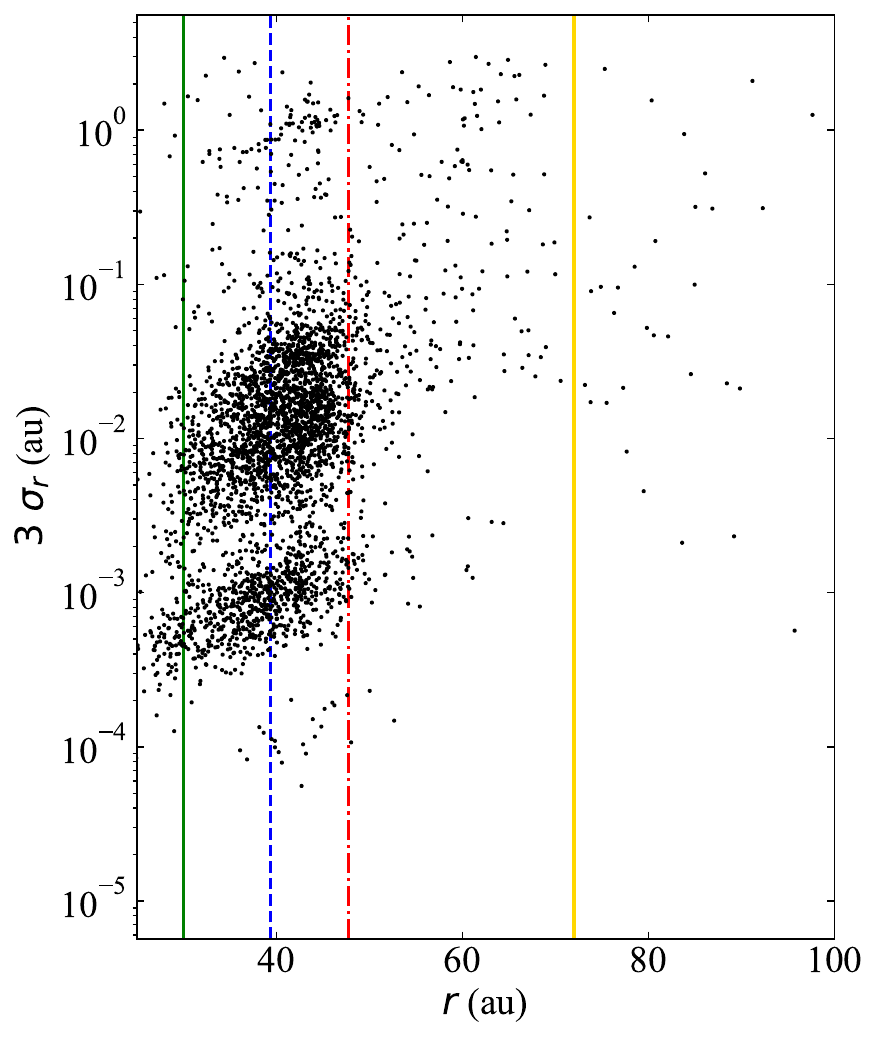}
        \includegraphics[width=0.33\linewidth]{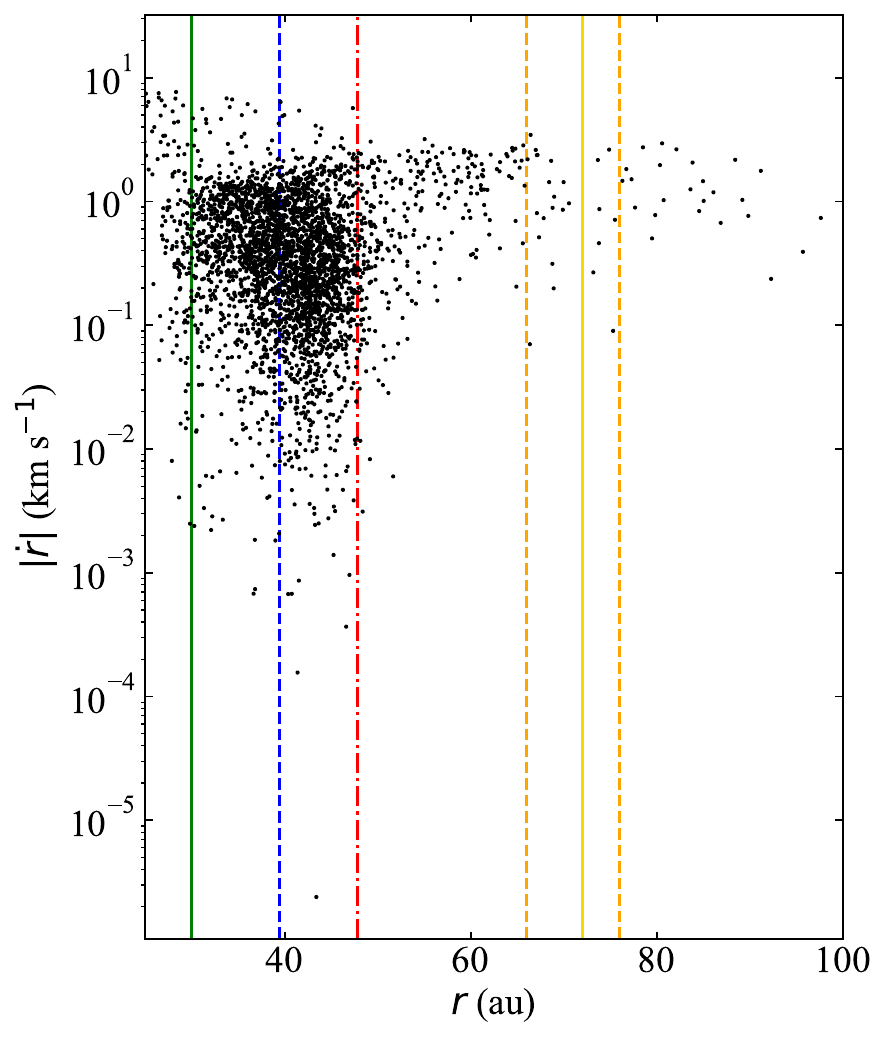}
        \includegraphics[width=0.33\linewidth]{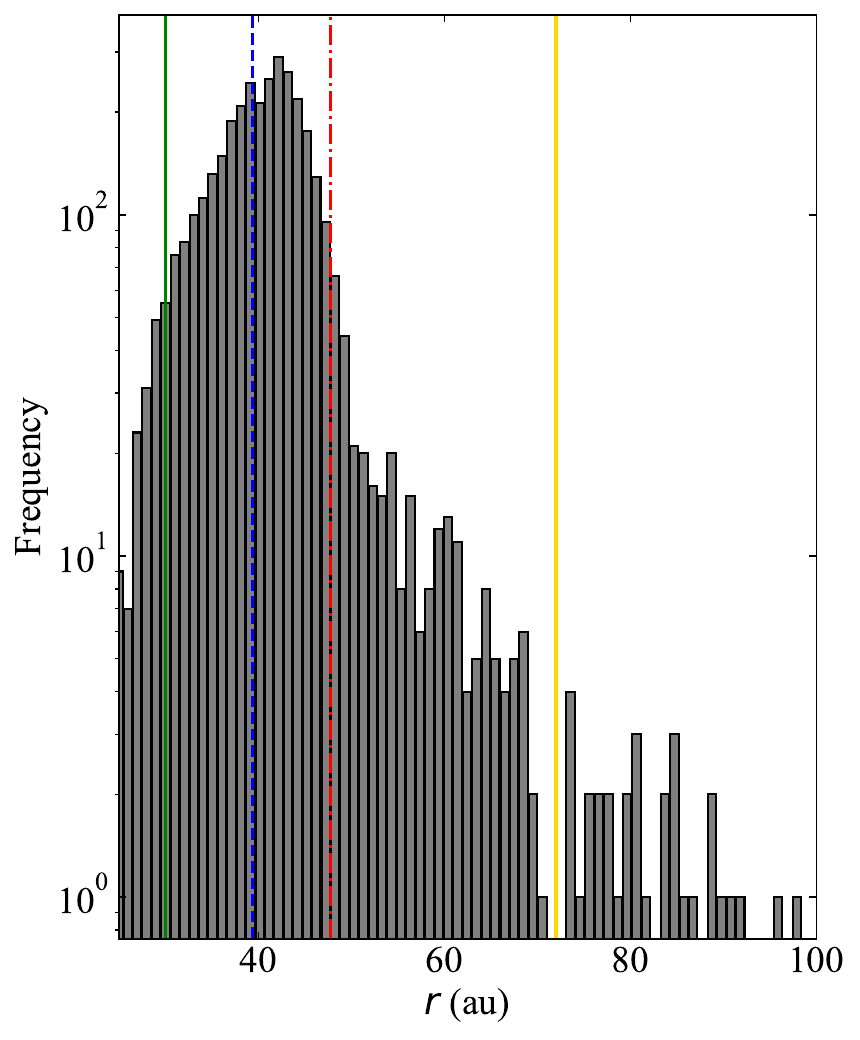}
        \caption{Heliocentric range and range-rate for the best sample ($\sigma_{r}<1$~au, 3595 objects). The left-hand 
                 side panel shows the uncertainty in heliocentric range as a function of the range. The central panel 
                 shows the absolute value of the heliocentric range-rate as a function of the range. The right-hand side 
                 panel shows the histogram of heliocentric ranges. The location of the 1:1 mean-motion resonance with 
                 Neptune at 30.0~au is displayed as a green solid vertical line, the 2:3 resonance at 39.4~au as blue 
                 dashed and the 1:2 resonance at 47.8~au as red dot-dashed. The golden vertical line is plotted at 72~au 
                 indicating the position of the gap discussed in the text. In the central panel, the two orange dashed 
                 vertical lines at 66~au and 76~au signal where $|\dot{r}|$ decreases. Data source: JPL's Horizons.
                }
        \label{uncertaintyB}
     \end{figure*}

     If we consider the subsample with $\sigma_{r}<1$~au that includes 3595 objects and repeat the analysis carried out 
     in Section~4, we obtain Fig.~\ref{uncertaintyB}. It shows a conspicuous gap around $r\sim$70~au (golden vertical 
     line at 72~au). Considering Poisson statistics ($\sigma$=$\sqrt{n}$) to compute uncertainties --- applying the 
     approximation given by \citet{1986ApJ...303..336G} when $n$$<$21, $\sigma$$\sim$$1+\sqrt{0.75+n}$ --- we obtain a 
     statistical significance close to 2$\sigma$ for this gap. The significance is similar for the full sample. 
     Figure~\ref{uncertaintyB}, central panel, shows that at both sides of the gap there is a trend for the absolute 
     value of the heliocentric range-rate to decrease (orange dashed vertical lines at 66~au and 76~au). Although the 
     statistical effect is only marginally significant, its overall properties are consistent with those of a real gap.  
%
%
%
%

\begin{figure}
       \centering
        \includegraphics[width=\linewidth]{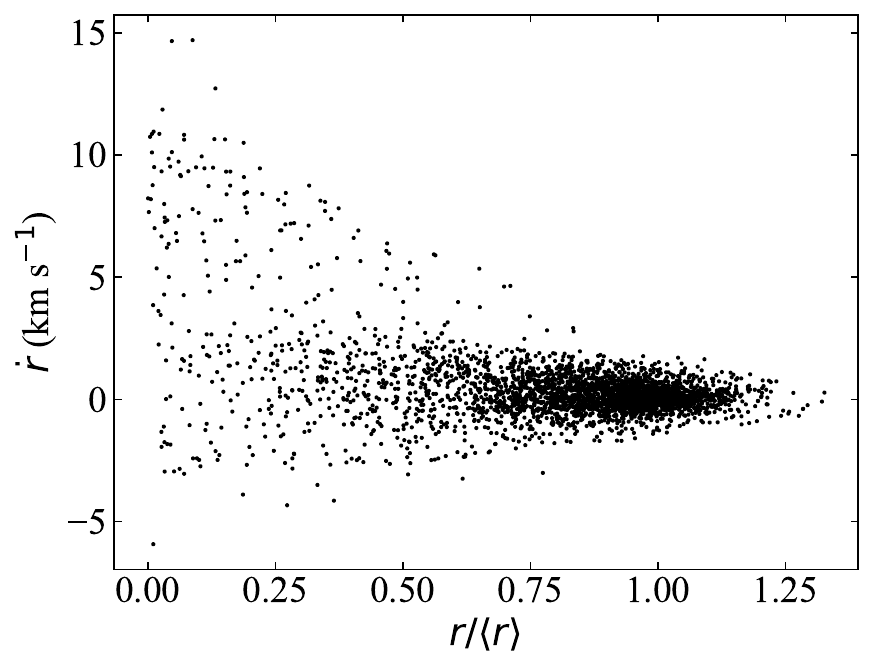}
        \caption{Heliocentric range-rate as a function of the radial distance normalized to $\langle{r}\rangle$ for the 
                 best sample ($\sigma_{r}<1$~au, 3595 objects). The plot shows that the sample is not strongly 
                 perihelion biased. Data source: JPL's Horizons.
                }
        \label{range-rate}
     \end{figure}

     For an elliptic orbit, the averaged distance over an orbital period is given by $\langle{r}\rangle=a\ (1+e^{2}/2)$. 
     If we represent the heliocentric range-rate of the best sample as a function of the range divided by 
     $\langle{r}\rangle$ (see Fig.~\ref{range-rate}), we observe that most objects are currently located at 
     $r\leq\langle{r}\rangle$ although a certain fraction is probably close to aphelion as their heliocentric 
     range-rates are small. In other words, the sample is not entirely biased towards objects at perihelion and probes
     the entire range under study. Unfortunately, nearly all these objects at aphelion are relatively recent discoveries 
     (2016 onwards) with short data arcs. The group with range-rate above $\sim$3~km~s$^{-1}$ is made of objects in 
     eccentric orbits observed near perihelion.
%
%
%
%

  \section{Discussion}
     The radial distribution of trans-Neptunian objects was studied before; for example, \citet{2001ApJ...554L..95T} 
     found an outer edge for the belt at 47$\pm$1~au but the data they analysed could not reject the presence of a 
     second belt beyond 76~au. In general terms, our results confirm theirs.
     The tail of the distribution in Fig.~\ref{uncertaintyB}, right-hand side panel, shows signs of orbital substructure
     that, in principle, may also be attributed to observational biases and/or lack of completeness. The only other 
     well-studied structure that may help in understanding the nature of the apparent gap at $r\sim$70~au is the outer 
     main asteroid belt that is shaped by Jupiter via close encounters, mean-motion and secular resonances, and by 
     secular resonances with Saturn. 

\begin{figure}
       \centering
        \includegraphics[width=\linewidth]{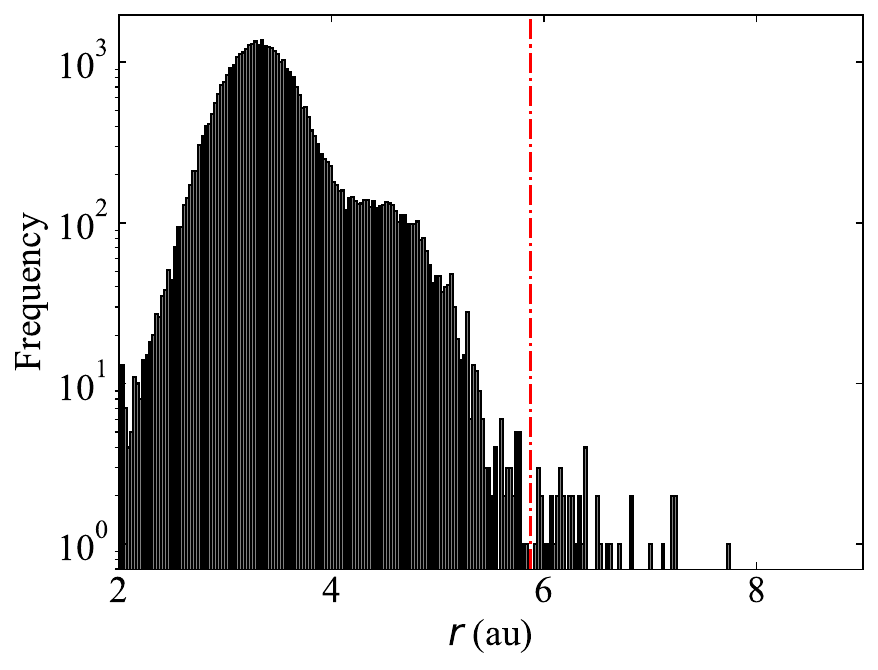}
        \caption{Histogram of heliocentric ranges for outer main-belt asteroids (39933 objects). The red dot-dashed 
                 vertical line signals the 5:6 mean-motion resonance with Jupiter at 5.875~au. Data source: JPL's 
                 Horizons.
                }
        \label{ombrange}
     \end{figure}

     Figure~\ref{ombrange} shows that the tails of the heliocentric range distributions of trans-Neptunian objects and 
     outer main-belt asteroids are not too dissimilar. Objects beyond the gap observed in Fig.~\ref{ombrange} close to 
     6~au --- red dot-dashed vertical line signalling the 5:6 mean-motion resonance with Jupiter at 5.875~au, see 
     panel~c in fig.~7 of \citet{2006Icar..184...29G} --- are asteroids with comet-like 
     orbits\footnote{\href{https://physics.ucf.edu/~yfernandez/lowtj.html}
     {https://physics.ucf.edu/$\sim$yfernandez/lowtj.html}} analogues to those of JFCs. Figure~\ref{ombrange} shows that 
     the Kuiper Cliff is not dramatically different from the feature observed in the outer main-belt. A separate issue 
     is the mechanism responsible for the observed tail of the distribution. As Jupiter is the main perturber of the 
     JFCs, Neptune might be responsible for the gap at $\sim$70~au, but panel~f in fig.~7 of \citet{2006Icar..184...29G}
     suggests otherwise. In addition, both 90377 Sedna (2003~VB$_{12}$) and 2012~VP$_{113}$, two widely regarded IOCOs, 
     have perihelia beyond the gap. Two additional objects with rather poor orbit determinations also have nominal 
     perihelia beyond the gap: 2019~EE$_{6}$, a retrograde ETNO, and 2020~MJ$_{53}$, a nominal low-eccentricity, 
     low-inclination object. Only these four have $q>70$~au.  
%
%
%
%

\begin{figure}
       \centering
        \includegraphics[width=\linewidth]{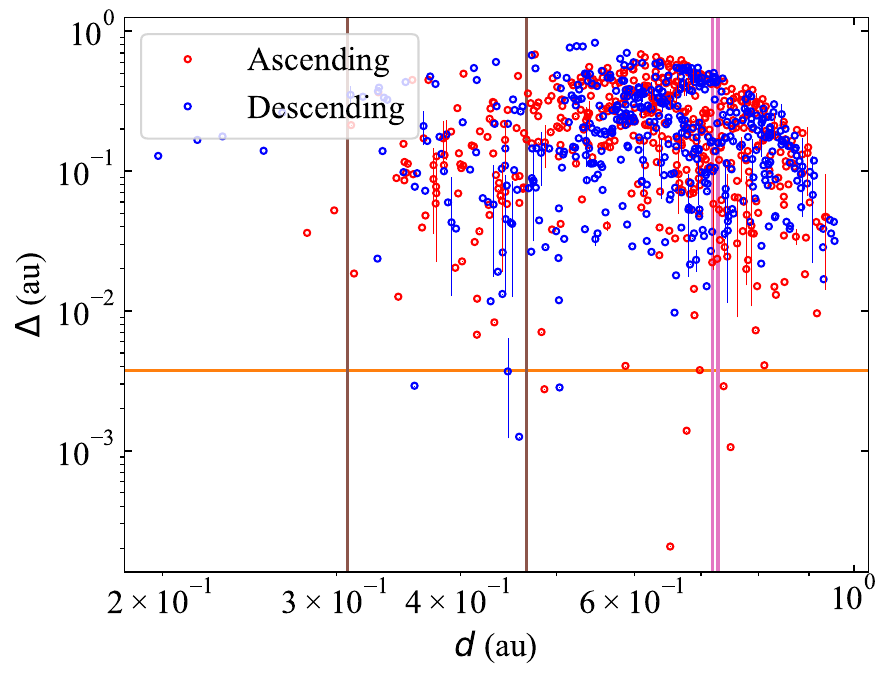}
        \caption{Mutual nodal distances as a function of the heliocentric distance to the node for the sample of 31 
                 Atiras or 465 pairs. The values of mutual nodal distances of ascending nodes, ${\Delta}_{+}$, are shown 
                 in red and those of descending nodes, ${\Delta}_{-}$, in blue. The orange horizontal line at 0.0037~au
                 signals the severe outlier boundary (1st percentile). The pink vertical bars show Venus' perihelion and
                 aphelion, the brown vertical bars display those of Mercury. Data source: JPL's SBDB.
                }
        \label{nodaldistatiras}
     \end{figure}

     Gaps can also be induced by a massive perturber located beyond the gap. Our result is somewhat consistent with the 
     conclusions in \citet{2021AJ....162...39O}. The existence of a moderately massive planet in the neighbourhood of, 
     but beyond, the feature identified by our analysis has been proposed multiple times (see e.g. 
     \citealt{2002Icar..160...32B,2002Icar..157..269G,2008AJ....135.1161L,2016MNRAS.455L.114W,2017AJ....154...62V,
     2023AJ....166..118L}). But  
     how strong is the evidence for the presence of such a perturber? An unexpected supporting argument comes from the
     analysis of the distribution of mutual nodal distances of the Atiras, a population of asteroids with aphelia inside
     the orbit of Earth. The mutual nodal distances of two objects can be computed using eqs.~16 and 17 in 
     \citet{2017CeMDA.129..329S}. A small mutual nodal distance may signal a pair of fragments of a larger body or the
     components of a disrupted binary. Close encounters with planets may lead to fragmentation events and binary 
     disruptions. Figure~\ref{nodaldistatiras} shows the distribution of mutual nodal distances for the 465 pairs 
     defined by the 31 known Atiras. The shortest nodal distances are strongly correlated with the regions occupied by
     Venus (pink vertical bars) and Mercury (in brown).
%
%
%
%

\begin{figure}
       \centering
        \includegraphics[width=\linewidth]{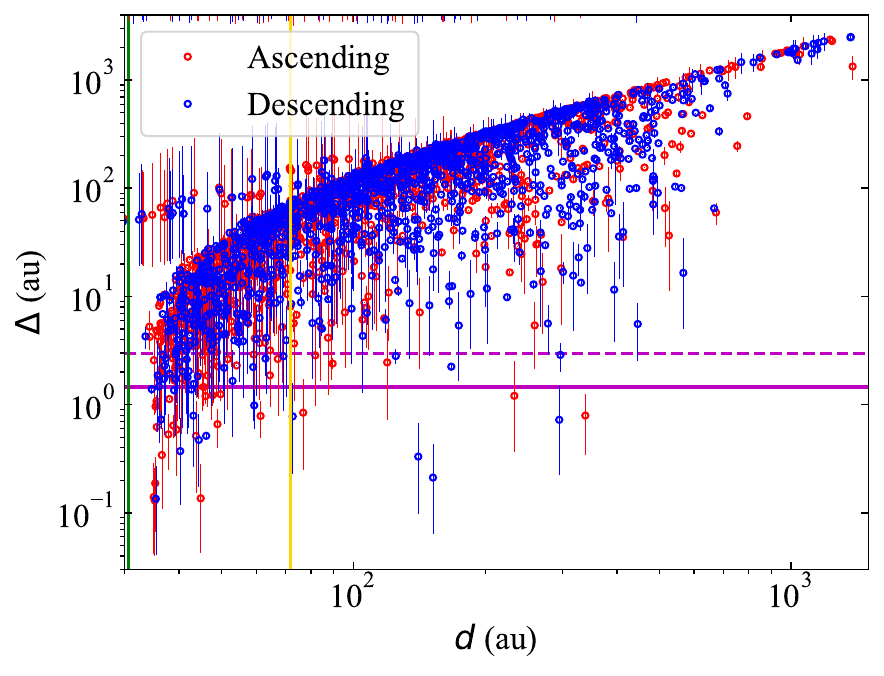}
        \caption{Mutual nodal distances as a function of the barycentric distance to the node for the sample of 56 ETNOs 
                 or 1540 pairs. The values of mutual nodal distances of ascending nodes, ${\Delta}_{+}$, are shown in 
                 red and those of descending nodes, ${\Delta}_{-}$, in blue. The solid purple line corresponds to 
                 1.455~au and the dashed one to 2.975~au that signal outliers inside the 1st percentile. The golden 
                 vertical line plotted at 72~au shows the position of the gap discussed in the text. Data source: JPL's 
                 SBDB.
                }
        \label{nodaldistetnos}
     \end{figure}

     Performing the same analysis for the ETNOs --- with perihelion distances, $q>30$~au and $a>150$~au as defined by 
     \citet{2014Natur.507..471T} --- we obtain Fig.~\ref{nodaldistetnos}, which is an updated version of the ones in 
     \citet{2021MNRAS.506..633D} and \citet{2022MNRAS.512L...6D}, and includes 56 objects and 1540 pairs. Neglecting the 
     short nodal distances close to perihelion, the shortest set of distances is found in the interval of barycentric 
     distances (100,~400)~au. Binaries are likely dominant beyond Neptune \citep{2017NatAs...1E..88F} and wide binaries 
     can be disrupted following encounters with a massive perturber \citep{2017Ap&SS.362..198D}. Observational evidence 
     of this process may have already been uncovered by \citet{2017MNRAS.467L..66D}. If the pairs with unusually short 
     mutual nodal distances are the result of a planetary encounter, Fig.~\ref{nodaldistetnos} signals the likely region 
     inhabited by the perturber. 
%
%
%
%

  \section{Conclusions}
     In this Letter, we applied a methodology that uses the heliocentric range and range-rate of the known Kuiper belt 
     objects and their uncertainties to identify structures beyond the Kuiper Cliff. This approach is more reliable than 
     those based on the orbital elements because the uncertainties in heliocentric range and range-rate are smaller than 
     those of orbital parameters like the semimajor axis. We show that the distribution in heliocentric range beyond the 
     Kuiper Cliff resembles that of the outer main asteroid belt and exhibits a gap at $\sim$70~au that might be caused 
     by Neptune but also by the combined action of external perturbers. The analysis of the outliers in the distribution 
     of mutual nodal distances of the ETNOs suggests that a massive perturber is present beyond the heliopause.

  \section*{Acknowledgements}
     We thank the anonymous referee for a constructive, helpful and timely report, S.~J. Aarseth, J. de Le\'on, J. 
     Licandro, A. Cabrera-Lavers, J.-M. Petit, M.~T. Bannister, D.~P. Whitmire, G. Carraro, E. Costa, D. Fabrycky, A.~V. 
     Tutukov, S. Mashchenko, S. Deen, and J. Higley for comments on ETNOs, and A.~I. G\'omez de Castro for providing 
     access to computing facilities. This work was partially supported by the Spanish `Agencia Estatal de 
     Investigaci\'on (Ministerio de Ciencia e Innovaci\'on)' under grant PID2020-116726RB-I00 
     /AEI/10.13039/501100011033. In preparation of this Letter, we made use of the NASA Astrophysics Data System, the 
     ASTRO-PH e-print server, and the MPC data server.

  \section*{Data Availability}
     The data underlying this paper were accessed from JPL's Horizons (\href{https://ssd.jpl.nasa.gov/?horizons}
     {https://ssd.jpl.nasa.gov/?horizons}). The derived data generated in this research will be shared on reasonable 
     request to the corresponding author.

  \bsp
  \label{lastpage}
\end{document}